\documentstyle[aps,prb]{revtex}
\def\eps{\varepsilon}
\def\del{\nabla}
\def\dd{\partial}
\def\ceffi{ \sigma_1}
\def\ceff{ \sigma_{\rm eff}}
\def\seff{ S_{\rm eff}}
\def\keff{ \kappa_{\rm eff}}
\def\ext{{\rm ext}}
\def\ind{{\rm ind}}

\def\gt{{\bf{\nabla}}T}
\def\usigma{\underline{\sigma}}
\def\ukappa{\underline{\kappa}}
\def\uS{\underline{S}}

\def\uA{\underline{A}}
\def\kk{{\bf k}}
\def\ts{\tilde{\sigma}_0}
\def\pmb#1{\setbox0=\hbox{#1}%
 \kern-.025em\copy0\kern-\wd0
 \kern.05em\copy0\kern-\wd0
 \kern-.025em\raise.0433em\box0}

\begin{document}

\draft
\preprint{Submitted to Phys. Rev. B}

\title{Thermoelectric Properties of Anisotropic Systems}
\author{W. E. Bies, R. J. Radtke, and H. Ehrenreich}
\address{Physics Department and
the Division of Engineering and Applied Sciences,\\
Harvard University, Cambridge, Massachusetts~~02138}
\maketitle

\begin{abstract}
The effective transport coefficients and figure of merit $ZT$
for anisotropic systems are derived from a macroscopic formalism.
The full tensorial structure of the transport coefficients
and the effect of the sample boundaries are included.
Induced transverse fields develop which can be larger than the applied
fields and which reduce the effective transport coefficients.
A microscopic model relevant for multi-valleyed materials is introduced
which utilizes the effective-mass and relaxation-time approximations.
The thermopower and Lorentz number are independent of the tensorial
structure of the transport coefficients in this case and are therefore
isotropic.
$ZT$ is also isotropic for vanishing lattice thermal conductivity
$\kappa_\ell$.
For non-vanishing but sufficiently isotropic $\kappa_\ell$, $ZT$ is maximal
along the direction of highest electrical conductivity $\sigma$.
Numerical calculations suggest that maximal $ZT$ generally occurs along
the principal direction with the largest $\sigma/\kappa_\ell$.
An explicit bound on $ZT$ is derived.
Several results for specific systems are obtained:
(1) Bulk n-type $\rm Bi_2Te_3$ exhibits easily observable
induced transverse fields and anisotropic $ZT$s.
(2) Increased anisotropy in HgTe/$\rm Hg_{1-x}Cd_xTe$ superlattices
(SLs) is associated with larger induced fields.
(3) The valley degeneracy is split and the bulk masses modified
in isolated $\rm Bi_2Te_3$ quantum wells, resulting in optimal $ZT$s
for wells grown along the trigonal direction.
(4) Non-parabolic dispersion in SLs has little effect on the thermopower
at the carrier concentrations which maximize $ZT$.
\end{abstract}

\pacs{PACS numbers:  72.20.Pa, 72.10.Bg, 73.20.Dx}


\section{Introduction}

In the search to find systems with large thermoelectric figures of merit
$ZT$, the emphasis has been more on new materials than on materials
structures or crystallographic anisotropy.
Typical of new structures are superlattices, quantum
wells\cite{Hicks2D,MahanQW,SofoSL,Broido95,Hicks96,Broido97,HicksBi,BroidoBi}
and quantum wires.\cite{Hicks1D,Broido1D}
It has been generally assumed that the direction of highest conductivity
in an anisotropic material yields optimal thermoelectric properties.
The correctness of this assumption is not obvious, since directions
may exist along which the lattice thermal conductivity is abnormally
low and the thermopower is high enough to result in large $ZT$ even
though the electrical conductivity is less than maximum.

This paper develops a more general transport theory of thermoelectricity
in anisotropic systems.
The ``highest conductivity'' assumption is found to be correct for materials
having simple band structures typically of the parabolic variety and
essentially isotropic lattice thermal conductivities.
However, the formalism developed here suggests more generally that
the optimal orientation corresponds to the principal direction along
which the ratio of the electronic to lattice thermal conductivity
$\sigma/\kappa_\ell$ is maximum.
There are also several surprising results.
These include the formation of possibly large induced transverse
electric fields and temperature gradients, the fact that $ZT$ is strictly
isotropic in anisotropic systems having parabolic bands if the
lattice thermal conductivity is neglected, and that a nearly isotropic
thermopower and Lorentz number results under these conditions even if
the bands are extremely anisotropic and non-parabolic.

The macroscopic formalism, based on the tensorial form of the usual
phenomenological transport equations, is presented in Sec.~\ref{sec:formalism}.
The effects of sample boundaries are included by requiring that the
transverse electric and heat currents vanish at the transverse surfaces.
Anisotropic and isotropic systems are shown to differ both
qualitatively, through the presence of induced transverse fields, and
quantitatively, through the magnitude of the transport coefficients.

More detailed statements concerning optimal orientations require use of
a microscopic model.
Section~\ref{sec:micro} introduces a model commonly used to
study transport in semiconductors having multi-valley,
anisotropic band structures.
The transport properties follow from the linearized Boltzmann equation
in the effective-mass and relaxation-time approximations.
Intervalley scattering is neglected.
The thermopower and Lorentz number somewhat surprisingly turn out to be
isotropic.
The same is true for $ZT$ when the lattice thermal conductivity
$\kappa_\ell$ is neglected.
Sec.~\ref{sec:lowD} shows that similar results hold for two- and
one-dimensional systems.
The realistic case corresponding to finite $\kappa_\ell$ is considered
in Sec.~\ref{sec:bound}.
If $\kappa_\ell$ is sufficiently isotropic, the maximal $ZT$ is shown
to occur for samples cut along the direction of highest electrical
conductivity.
This rather unsurprising result, however, may be modified if the anisotropy
of $\kappa_\ell$ exceeds that of the electrical conductivity,
leading to the conjecture concerning $\sigma/\kappa_\ell$ mentioned above.
An explicit expression for the upper bound on $ZT$ is also derived,
which is a generalization of that previously obtained\cite{Mahan} to
anisotropic systems.

This microscopic description is then applied to several materials of
current interest.
Sec.~\ref{sec:BiTeBulk} examines bulk n-type $\rm Bi_2Te_3$, in particular,
the magnitude of the induced fields, their influence on the effective
electrical conductivity, and the dependence of $ZT$ on sample orientation.
HgTe/HgCdTe superlattices (SLs), considered in Sec.~\ref{sec:sl}, provide
an example of systems which have a tunable anisotropy.
For sufficiently large anisotropy, the induced fields can be much larger
than the external one.
Sec.~\ref{sec:BiTeQW} focuses on isolated n-type $\rm Bi_2Te_3$
quantum wells.
Quantum confinement is shown to split the valley degeneracy
and modify the effective masses relative to their bulk values.
These effects are seen to be important in determining $ZT$ quantitatively.

The effects of non-parabolicity in superlattices on the thermopower and
Lorentz number are investigated in Sec.~\ref{sec:anisS}.
Both quantities are seen to be nearly isotropic.
The thermopower is also bounded by its values along the principal axes,
supporting the conjecture that optimal $ZT$ is obtained along those
directions.

\section{Electronic Transport Theory}
\label{sec:formalism}

Consider a rectangular parallelepiped sample of an anisotropic material which
is placed between a hot contact at {$x=0$}, temperature {$T_h$}, and a cold 
contact at {$x=L_x$} of temperature {$T_c$}.
An external electric field
${\bf \cal E}^{\rm ext}=({\cal E}^{\rm ext}_\|,0,0)$ is applied to drive a
current from the hotter to the colder electrode.
Because of the anisotropy, the resulting electric current need not be parallel
to the {$x$}-direction initially; instead it may have a transverse component
that leads to an accumulation of charge on the faces of the sample, resulting
in an induced polarization field.
Similarly, the initial thermal current through the sample need not be
parallel to $x$, and this may lead to an induced transverse temperature
gradient.
In the following, we compute the effect of these induced fields and
temperature gradients on the electrical conductivity, thermopower,
thermal conductivity, and $ZT$ using phenomenological linear-response theory.
We assume that the transverse electrical and thermal currents vanish.

The electric current is
\begin{equation}
{\bf J}^e = \usigma ({\bf {\cal E}}- \uS \gt) \equiv \usigma {\bf F}
\label{1}
\end{equation}
where {$\usigma$} and {$\uS$} are the conductivity and Seebeck
tensors respectively.
Write ${\bf F} = (F_x,F_y,F_z) = (F_\|,{\bf F}_\bot)$ with $F_\|$ the
thermoelectric field along $x$ and ${\bf F}_\bot$ the induced transverse
fields.
The conductivity tensor takes the form
\begin{equation}
\usigma = \pmatrix{ \sigma_\parallel & \usigma_{od}^T \cr
{\usigma_{od}} & \usigma_\perp \cr},
\label{9}
\end{equation}
where {$\sigma_\parallel$} is the component of the conductivity along the
{$x$}-direction, and {$\usigma_\perp$} is a {$2 \times 2$} tensor for the
transverse {$y,z$}-directions.
Since the full conductivity tensor is not block-diagonal, there will be a
{$2 \times 1$} off-diagonal term {$\usigma_{od}$} and its transpose.

Imposing the boundary condition {${\bf J}^e=(J_\|^e,0,0)$} on
Eq.~(\ref{1}) corresponding to vanishing transverse current gives
\begin{equation}
\pmatrix{J_\|^e \cr {\bf 0} \cr}
= \usigma \pmatrix{ F_\parallel \cr {\bf F}_\perp \cr}.
\label{11}
\end{equation}
Inserting Eq.~(\ref{9}) into this expression leads to a linear equation
for ${\bf F}_\perp$ with solution
\begin{equation}
{\bf F}_\perp = - {\usigma_\perp}^{-1} {\usigma_{od}} F_\parallel.
\label{13}
\end{equation}
The off-diagonal elements of {$\usigma$} induce a current
\begin{equation}
J_\|^{\rm e,ind} = \usigma_{od}^T {\bf F}_\bot
  = -\usigma_{od}^T {\usigma_\perp}^{-1} {\usigma_{od}} F_\parallel
\label{6}
\end{equation}
along the $x$-direction.
Since $\usigma_\bot$ and $\usigma_\bot^{-1}$ have positive eigenvalues,
the quadratic form $\usigma_{od}^T {\usigma_\perp}^{-1} {\usigma_{od}}$
is positive definite.
Thus, the induced current opposes the external current
$J_\|^{\rm e,ext} = \sigma_\| F_\|$.
The total current
\begin{equation}
J_\|^e = J_\|^{\rm e,ext} + J_\|^{\rm e,ind} = \sigma_1 F_\|
\label{eq:Jparallel}
\end{equation}
with
\begin{eqnarray}
\sigma_1 &=& \sigma_\parallel - \usigma_{od}^T
{\usigma_\perp}^{-1} {\usigma_{od}} \label{eq:sigma1} \\
&=& \sigma_{xx} -
\sigma_{xy} {{\sigma_{yx}\sigma_{zz}-\sigma_{yz}\sigma_{zx}} \over
             {\sigma_{yy}\sigma_{zz}-\sigma_{yz}\sigma_{zy}}} -
\sigma_{xz} {{\sigma_{zx}\sigma_{yy}-\sigma_{zy}\sigma_{yx}} \over
             {\sigma_{yy}\sigma_{zz}-\sigma_{yz}\sigma_{zy}}} .
\label{14}
\end{eqnarray}
The induced fields therefore lead to a reduced conductivity:
{$\ceffi \le \sigma_\parallel$}.
If $\uS$ is diagonal, $\sigma_1$ is the effective conductivity; otherwise,
the induced temperature gradients give rise to additional terms considered
below.
In the isotropic case, the induced fields vanish, and {$\ceffi=\sigma_\|$}.
Note that in general $\ceffi \ge 0$ because
\begin{equation}
\ceffi F_\parallel^2 = J_\|^e F_\parallel = {\bf F}^T \cdot {\bf J}^e
= {\bf F}^T \cdot \usigma \cdot {\bf F} \ge 0.
\label{15c}
\end{equation}

The heat current {${\bf J}^Q$} is given by
\begin{equation}
{\bf J}^Q = T \uS {\bf J}^e - \ukappa \gt,
\label{15}
\end{equation}
where {$\ukappa$} is the thermal conductivity and the temperature
gradient $\gt = (\del_x T,\del_y T,\del_z T) =
(\del_\| T, {\bf \del}_\bot T)$.
Using the same decomposition as in Eq.~(\ref{9}) for {$\ukappa$} and
{$\uS$} in Eq.~(\ref{15}) and setting the transverse components of the heat
current to zero yields
\begin{equation}
\pmatrix{ J^Q_\| \cr {\bf 0} \cr} = 
\pmatrix{ T S_\parallel J_\|^e - \kappa_\parallel \del_\| T
- \ukappa_{od}^T {\bf \del}_\bot T \cr T \uS_{od} J_\|^e - \ukappa_{od}
\del_\| T - \ukappa_\perp {\bf \del}_\bot T \cr} .
\label{15a}
\end{equation}
Solving for the induced temperature gradients yields
\begin{equation}
{\bf \del}_\bot T = \ukappa_\perp^{-1} (T \uS_{od} J_\|^e - 
\ukappa_{od} \del_\| T ).
\label{15b}
\end{equation}
Substituting back into Eq. (\ref{15a}) gives
\begin{equation}
J_\|^Q = T\seff J_\|^e - \keff \del_\| T
\label{20} 
\end{equation}
with
\begin{eqnarray}
\keff &=& \kappa_\parallel - \ukappa_{od}^T \ukappa_\perp^{-1} \ukappa_{od}
\label{21a} \\
\seff &=& S_\parallel - \ukappa_{od}^T \ukappa_\perp^{-1} \uS_{od}.
\label{21}
\end{eqnarray}
Observe that {$\ceffi$}, Eq. (\ref{eq:sigma1}), and {$\keff$}, Eq. (\ref{21a}),
have the same form.
As in the $\sigma_1$ case, the induced temperature gradients
produce an induced heat current along $x$ which opposes the external
heat current $-\kappa_\| \del_\| T$.
The net result is a reduced effective thermal conductivity:
$0 \le \keff \le \kappa_\|$.

Returning to Eq.~(\ref{eq:Jparallel}), we have
\begin{equation}
J_\|^e = \ceffi F_\parallel =
\ceffi \left({\cal E}_\parallel - S_\parallel \del_\| T
- \uS_{od}^T {\bf \del}_\bot T \right).
\label{23}
\end{equation}
Substituting Eq.~(\ref{15b}) into Eq.~(\ref{eq:Jparallel}) leads to
\begin{equation}
J_\|^e = \ceff( {\cal E}_\parallel - \seff \del T_\parallel).
\label{24}
\end{equation}
with
\begin{equation}
\ceff = {\ceffi \over {1+\ceffi T \uS_{od}^T \ukappa_\perp^{-1} \uS_{od}}}
\label{24a}
\end{equation}
and
\begin{equation}
\seff = S_\parallel - \uS_{od}^T \ukappa_\perp^{-1} \ukappa_{od}.
\label{24b}
\end{equation}
From the properties of $\sigma_1$, $0 \le \ceff \le \sigma_\|$.
Therefore, induced fields and temperature gradients always reduce
the effective conductivity.
Also, {$\seff$} in Eq. (\ref{24b}) is the same as $\seff$ in Eq.~(\ref{21})
because, by the Onsager relations, {$\ukappa_\perp$} and hence 
{$\ukappa_\perp^{-1}$} are symmetric. 

In the steady state, the energy-conservation equation reads
\begin{equation}
\del \cdot {\bf J}^Q = {\cal E} \cdot {\bf J}^e.
\label{24c}
\end{equation}
Using Eqs.~(\ref{15}) and (\ref{1}), and assuming that the Thompson heat
is negligible,
\begin{equation}
\del \cdot {\bf J}^Q = \del \cdot ( T \uS {\bf J}^e - \ukappa {\bf \del} T )
= {\bf \del} T \uS {\bf J}^e - {\bf \del} \cdot
( \ukappa {\bf \del} T)
\end{equation}
and
\begin{equation}
{\cal E} \cdot {\bf J}^e = {\bf J}^e ( \usigma^{-1} {\bf J}^e
  + \uS {\bf \del} T)
= {\bf J}^e \usigma^{-1} {\bf J}^e + {\bf J}^e \uS {\bf \del} T.
\end{equation}
Thus, Eq. (\ref{24c}) becomes
\begin{equation}
{\bf J}^e \usigma^{-1} {\bf J}^e + \del \cdot ( \ukappa {\bf \del} T) = 0.
\end{equation}
Since ${\bf J}^e = (J_\|^e,0,0)$, the first term is
{$(\usigma^{-1})_{11}J_\|^{e2}=(\det \usigma_\perp/\det \usigma)
J_\|^{e2}=\sigma_1^{-1} J_\|^{e2}$}.
The last step is justified by expanding the determinants and comparing
to Eq.~(\ref{14}).
With ${\bf \del} T 
= (\del_\| T, {\bf \del}_\bot T)$ and using Eq.~(\ref{15b}), the second
term is 
\begin{equation}
{\bf \del} \cdot
\pmatrix{ \kappa_\| \del_\| T + \ukappa_{od}^T {\bf \del}_\bot T \cr
\ukappa_{od} \del_\| T + \ukappa_\perp {\bf \del}_\bot T \cr}
= {\bf \del} \cdot
\pmatrix{ (\kappa_\parallel - \ukappa_{od}^T \ukappa_\perp^{-1}
\ukappa_{od}) \del_\| T + \ukappa_{od}^T \ukappa_\perp^{-1} \uS_{od}
T J_\|^e \cr T \uS_{od} J_\|^e \cr}
\end{equation}
\begin{equation} = \keff \del_\|^2 T + \uS_{od}^T \ukappa_\perp^{-1}
\uS_{od} T J_\|^{e2}.
\end{equation}
Therefore Eq. (\ref{24c}) reduces to
\begin{equation}
\ceff^{-1} J_\parallel^{e2} + \keff \del_\|^2 T = 0.
\label{25a}
\end{equation}

With the help of Eqs.~(\ref{24}), (\ref{20}) and (\ref{25a}), the
thermoelectric figure of merit becomes
\begin{equation}
ZT = T \ceff \seff^2/\keff .
\label{25}
\end{equation}
The transport coefficients are simply replaced by their effective versions.

These results are derived more succinctly using a general formalism
based directly on the Onsager coefficients in Appendix~A.
Rigorous bounds can be placed on the magnitude and sign of
$\sigma_{\rm eff}$ and $\kappa_{\rm eff}$, but not on $S_{\rm eff}$
without introducing a microscopic model that relates these transport
coefficients.
The next Section considers an example of such a model.

\section{Microscopic Model:
Parabolic Bands}
\label{sec:micro}

\subsection{Three-Dimensional Structures
\label{sec:3D}}

According to semiclassical
transport theory, the Boltzmann equation in the relaxation-time approximation
yields the transport coefficients
\begin{eqnarray}
\sigma_{ij} &=& e^2 \int d\eps ( - {{\dd f_0} / {\dd \eps}} ) 
\Sigma_{ij}(\eps) \label{26prime} \\
T(\sigma \cdot S)_{ij} &=& e \int d\eps ( - {{\dd f_0} / {\dd \eps}})
\Sigma_{ij}(\eps) (\eps-\mu) \\
T\kappa_{0,ij} &=& \int d\eps ( - {{\dd f_0} / {\dd \eps}} ) 
\Sigma_{ij}(\eps) (\eps-\mu)^2 ,
\label{26}
\end{eqnarray}
where {$f_0$} is the Fermi-Dirac
distribution {$f_0(\eps)=1/(\exp((\eps-\mu)/k_BT)+1)$},
{$\mu$} the chemical potential, and
\begin{equation}
\Sigma_{ij}(\eps) = \int {{2 d^3\kk} \over {(2\pi)^3}}
v_i(\kk) v_j(\kk) \tau(\kk) \delta(\eps-\eps(\kk))
\label{27}
\end{equation}
are the components of the transport distribution tensor, the generalization
of the function discussed by Mahan and Sofo.\cite{Mahan}
Here {$\eps(\kk)$} is the electronic dispersion relation,
{$v_i(\kk) = \hbar^{-1} {{\dd \eps(\kk)} / {\dd k_i}}$}
the electronic group velocity, and {$\tau(\kk)$} the relaxation time.
Note that {$\ukappa_0$} is the
electronic thermal conductivity at zero electrochemical potential
gradient inside the sample; \cite{Mahan}
the usual electronic thermal conductivity
at zero electric current, {$\ukappa_e$}, is given in terms of {$\ukappa_0$}
by {$\ukappa_e = \ukappa_0 - T \usigma \uS^2$}.

The microscopic model to be used here assumes the conduction to be taking
place in a single parabolic band having $N$ degenerate valleys centered at
{$\kk^{(n)}, n=1,\ldots,N$} respectively. The dispersion relation 
for each valley is
\begin{equation}
\eps^{(n)}(\kk) = \eps_0 + (\hbar^2/2) \sum_{i,j}
(k_i-k^{(n)}_i) M^{(n)-1}_{ij} (k_j-k^{(n)}_j) 
\label{30}
\end{equation}
with $\underline{M}^{(n)-1}$ the inverse effective-mass tensor.
The corresponding group velocity is
\begin{equation}
v_i^{(n)}(\kk) = \hbar^{-1} {{\dd \eps(\kk)} / {\dd k_i}}
= \hbar \sum_j M^{(n)-1}_{ij} (k_j - k^{(n)}_j).
\label{31}
\end{equation}

Intervalley scattering will be neglected. Thus the transport distribution 
tensor involves just a sum over the {$N$} valleys. 
Assuming the relaxation time to be a function of energy alone,
$\tau(\kk)=\tau(\eps(\kk))$, and independent of crystal orientation,
\begin{eqnarray}
\Sigma_{ij}(\eps) &=&
  \tau(\eps) \sum_{n=1}^N \int {{2 d^3\kk} \over {(2\pi)^3}}
  \hbar^2 \sum_{i',j'} M^{(n)-1}_{ii'} k_{i'} M^{(n)-1}_{jj'} k_{j'}
  \delta(\eps-\eps(\kk+ \kk^{(n)}))
  \label{32} \\
&=& \tau(\eps) \hbar^2 \sum_{n=1}^N \sum_{i'j'}
  M_{ii'}^{(n)-1}M_{jj'}^{(n)-1} \int k_{i'} k_{j'}
  \delta(\eps-{\bf k}\underline{X}^{(n)}{\bf k})
  {{2d^3{\bf k}} \over {(2\pi)^3}} \\
&=& {{2\tau(\eps)\hbar^2} \over {(2\pi)^3}} \sum_{n=1}^N \sum_{i'j'}
  M_{ii'}^{(n)-1}M_{jj'}^{(n)-1} \left[
  -{\partial \over {\partial X^{(n)}_{i'j'}}}
  \int \Theta(\eps-{\bf k}\underline{X}^{(n)}{\bf k}) d^3{\bf k} \right]
  \label{32a} \\
&=& {{2^{3/2}\tau(\eps)\eps^{3/2}} \over {3 \pi^2 \hbar^3}}
  \sum_{n=1}^N \sqrt{ \det \underline{M}^{(n)}} M^{(n)-1}_{ij},
\end{eqnarray}
where $X^{(n)}_{ij}$ are the components of
$\underline{X}^{(n)}=(\hbar^2/2)\underline{M}^{(n)-1}$.
This transformation relies explicitly on the validity of the effective-mass
approximation, the neglect of intervalley scattering, and the relaxation-time
approximation used here.
The square bracket in Eq.~(\ref{32a}) is evaluated using the identity
\begin{equation}
{\partial \over {\partial X^{(n)}_{ij}}} \det \underline{X}^{(n)} =
({\rm det} \underline{X}^{(n)}) X^{(n)-1}_{ij}
\end{equation}
and the change of variable ${\bf k'} = \sqrt{\underline{X}^{(n)}} {\bf k}$.
Thus,
\begin{equation}
\underline{\Sigma}(\eps)=\underline{A} {\cal T}(\eps)
\label{41}
\end{equation}
with
\begin{equation}
\underline{A} = \sum_{n=1}^N \left( m_0^{-1/2}
 \sqrt{{\rm det} \underline{M}^{(n)}} \right) \underline{M}^{(n)-1}
\label{39}
\end{equation}
and
\begin{equation}
{\cal T}(\eps) = {{2^{3/2} m_0^{1/2}} \over {3\pi^2\hbar^3}} \eps^{3/2} 
\tau(\eps).
\label{40}
\end{equation}
The constant, dimensionless matrix {$\uA$} contains the full tensorial
structure and separates it from the energy dependence in ${\cal T}(\eps)$.

The conductivity then becomes
\begin{equation}
\usigma = 
e^2 \int d\eps ( - {{\dd f_0} / {\dd \eps}} ){\cal T}(\eps)
\uA \equiv \sigma_0 \uA.
\label{44}
\end{equation}
Since
\begin{equation}
\underline{(\sigma \cdot S)} = (e/T) \int d\eps 
( - {{\dd f_0} / {\dd \eps}} ) {\cal T}(\eps)(\eps-\mu) \uA
\equiv \uA \sigma_0 S_0,
\label{46}
\end{equation}
\begin{equation}
\uS = \usigma^{-1} (\underline{\sigma \cdot S}) = S_0 \underline{1}.
\label{48}
\end{equation}
The Seebeck tensor is therefore necessarily {\it isotropic}.
Further,
\begin{equation}
\ukappa_0 = \uA \kappa_0
\label{49}
\end{equation}
with
\begin{equation}
\kappa_0 = T^{-1} \int d\eps ( - {{\dd f_0} / {\dd \eps}} ) 
{\cal T}(\eps) (\eps-\mu)^2.
\label{50}
\end{equation}
The zero-current electronic thermal conductivity becomes
\begin{equation}
\ukappa_e = \ukappa_0 - T \usigma \uS^2 \equiv \uA \kappa_e
\label{51}
\end{equation}
for $\kappa_e=\kappa_0-T\sigma_0 S_0^2$.

These results lead to the following surprising conclusion:
When the lattice thermal conductivity is neglected, then, within the
effective-mass approximation as specified here, the thermoelectric figure
of merit $ZT$ is independent of the sample orientation.
Note that
\begin{equation}
\ukappa_e \cdot \usigma^{-1} = \uA \kappa_e \cdot \uA^{-1} / \sigma_0
= (\kappa_e / \sigma_0) \underline{1} \equiv L_0T \underline{1},
\label{53}
\end{equation}
where the Lorentz
number {$L_0=\kappa_e/\sigma_0T$}. Thus,
{$\kappa_{e,ij} = L_0T\sigma_{ij}$} and  
\begin{equation}
\keff = L_0T\ceff
\label{54}
\end{equation}
as expected. Finally, since {$\uS$} is isotropic,
{$\seff = S_0$}. Combining these effective transport coefficients yields
\begin{equation}
ZT = T \ceff \seff^2 / \keff = S_0^2 / L_0,
\label{55}
\end{equation}
a constant independent of direction.

\subsection{Lower-dimensional Structures
\label{sec:lowD}}

Dimensionality enters
the transport coefficients through the {$\kk$}-space integrals
\begin{equation}
d^3k \qquad \leftrightarrow \qquad (2\pi/L_z) d^2k \qquad
\leftrightarrow \qquad ({{(2\pi)^2} / {L_yL_z}}) dk
\label{58}
\end{equation}
for three, two or one dimensions respectively, 
where {$L_y$} and {$L_z$} are the sample sizes in the
{$y$}- and {$z$}-directions.
Dimensionality also enters through the confinement
energies and effective masses for carriers constrained to move in a 
lower-dimensional device.

For the two-dimensional case with
${\bf \cal E} = ({\cal E}^\ext_x,{\cal E}^\ind_y)$ and
${\bf \del}T=(\del_xT,\del_y T)$, the induced fields are given by
Eqs.~(\ref{13}) and (\ref{15b}), the transport coefficients by
Eqs.~(\ref{21a}), (\ref{21}), and (\ref{24a}) with $\usigma_\perp$
replaced by $\sigma_{yy}$, $\usigma_{od}$ replaced by $\sigma_{yx}$;
similarly for the other transport coefficients.
The analogue of Eq.~(\ref{27}) for the components of the transport
distribution tensor is obtained within the effective-mass approximation
using Eq.~(\ref{58}):
\begin{eqnarray}
\Sigma_{ij}(\eps) &=& \frac{2\tau(\eps)\hbar^2}{4\pi^2L_z}
  \sum_{n=1}^N \sum_{i'j'} M^{(n)-1}_{ii'} M^{(n)-1}_{jj'} 
  \int {d^2\kk} k_{i'} k_{j'}
  \delta(\eps-{\bf k}\underline{X}^{(n)}{\bf k}).
\label{69}
\end{eqnarray}
This expression has the same form as Eq.~(\ref{41}) with
\begin{equation}
\underline{A} = \sum_{n=1}^N \sqrt{{\rm det} \underline{M}^{(n)}}
\underline{M}^{(n)-1}
\label{72}
\end{equation}
and
\begin{equation}
{\cal T}(\eps) = \eps \tau(\eps)/\pi \hbar^2 L_z.
\label{73}
\end{equation}
These equations are to be contrasted with Eqs. (\ref{39}) and (\ref{40})
for the three-dimensional case.
Thus, analogously to Eqs.~(\ref{44}), (\ref{48}), and (\ref{51}),
{$\usigma = \uA \sigma_0, \uS = S_0 \underline{1}$} and
{$\ukappa_e=\uA \kappa_e$}
where the two-dimensional {$\uA$} and {${\cal T}(\eps)$} 
must be used in defining {$\sigma_0$}, {$S_0$} and {$\kappa_0$}. Finally,
{$\ukappa_e \usigma^{-1}= L_0T \underline{1}$} so that
\begin{equation}
\keff = L_0 T\ceff.
\label{74}
\end{equation}
Also, {$\seff = S_0$} because {$\uS$} is
isotropic in two dimensions as well. The corresponding figure of merit
\begin{equation}
ZT = T \ceff \seff^2 / \keff = S_0^2 / L_0 
\label{75}
\end{equation}
is again independent of direction, as in three dimensions.

It is seen that the two-dimensional and three-dimensional results are
entirely analogous and that the former are obtained from the latter by
taking the limit as one of the
effective masses tends to infinity. In this limit, the ellipsoidal surface
in {$\kk$}-space of constant energy becomes increasingly prolate, until it
reaches the edge of the Brillouin zone, after which it assumes a cylindrical
shape extending from {$-\pi/L_z$} to {$\pi/L_z$} upon further increase in
the effective-mass parameter.
Furthermore {$ZT$} in two dimensions can be enhanced due to the factor of 
{$1/L_z$} in the density of states,
which, as pointed out in Ref.~\onlinecite{Hicks2D}, becomes large for small
thicknesses.

In the one-dimensional case there is no transport in the
transverse direction. Thus there are no transverse fields
and the microscopic and effective transport coefficients are the
same. Moreover, all transport coefficients are scalars. The transport
distribution function is found to be
\begin{equation}
\Sigma(\eps) =
  \sum_{n=1}^N {2 \over {\pi L_yL_z}}
  \sqrt{{{2\eps} \over {\hbar^2 m_x^{(n)}}}} \tau(\eps).
\label{76}
\end{equation}
Just as in two dimensions, the {$1/L_yL_z$} factor
leads to an enhancement of the density of states and thus of {$ZT$}
for thin wires.\cite{Hicks1D}
(The wire becomes approximately one-dimensional when it is thin enough
that the confinement energy places all states associated with its
three-dimensional structure sufficiently high in energy that they do
not contribute.)

\subsection{Implications for $ZT$
\label{sec:bound}}

We shall now derive an upper bound for $ZT$ of the Sofo and Mahan
form\cite{Mahan} and show that the highest conductivity direction gives
optimal values for $ZT$.
The inclusion of an isotropic lattice thermal conductivity $\kappa_\ell$
causes $ZT$ to lose its isotropy.

The figure of merit including an isotropic lattice thermal conductivity
$\underline{\kappa}_\ell = \kappa_\ell \underline{1}$ may be written in
the form
\begin{equation}
ZT = T \ceff \seff^2 / \keff
\label{77}
\end{equation}
where {$\ceff$} and {$\seff$} are as in Sec.~\ref{sec:formalism}.
$\keff = \kappa_e^* + \kappa_\ell$ with
\begin{eqnarray}
\kappa_e^* & = \kappa_{e,xx} &- \kappa_{e,xy}
{{\kappa_{e,yx}(\kappa_{e,zz}+\kappa_\ell)-\kappa_{e,yz}\kappa_{e,zx}} \over
{(\kappa_{e,yy}+\kappa_\ell)(\kappa_{e,zz}+\kappa_\ell)-\kappa_{e,yz}
\kappa_{e,zy}}} \nonumber \\
&&- \kappa_{e,xz}
{{\kappa_{e,zx}(\kappa_{e,yy}+\kappa_\ell)-\kappa_{e,zy}\kappa_{e,yx}} \over
{(\kappa_{e,yy}+\kappa_\ell)(\kappa_{e,zz}+\kappa_\ell)-\kappa_{e,yz}
\kappa_{e,zy}}}.
\label{86}
\end{eqnarray}
defining the electronic thermal conductivity in the presence of the
non-vanishing $\kappa_\ell$ and the sample boundaries.
As shown in Appendix~B, the upper bound on $ZT$ is given by
\begin{equation}
ZT \le a_0 ( \kappa_0 / \kappa_\ell) .
\end{equation}
The dimensionless quantity $a_0$ is defined by Eq.~(\ref{79}).
In the isotropic case,\cite{Mahan} $a_0 = 1$; in the present case,
$a_0$ is of order unity.
The maximum value of $\xi$ defined in Ref.~\onlinecite{Mahan} and in
Appendix~B [Eq.~(\ref{eq:xi})] of unity is attained if and only if
${\cal T}(\eps)$ is proportional to a $\delta$-function.
In the more physical case ${\cal T}(\eps) \propto \eps^r$ with $r$ varying
between -0.5 and 2, numerical computations show that $\xi$ tends to 1 as
$\mu/k_BT \rightarrow -\infty$ and to zero as $\mu/k_BT \rightarrow \infty$.
For $\mu/k_BT=0$, $\xi$ ranges from 0.5 to 0.8.
Thus, the upper bound can be reached at the cost of going to low
carrier concentrations, whereas higher carrier concentrations imply
smaller $\xi$.
The optimum concentration lies somewhere in between.

We now show that, in the effective-mass, relaxation-time, no intervalley
scattering, and isotropic-thermal-conductivity approximations,
{$ZT$} is highest in the direction of maximum electrical conductivity.
In the anisotropic case,
{$\ceff$} and {$\keff$} have an angular dependence 
obtained in the
frame of the sample by a rotation from their common principal frame.
{$\seff$} does not because in our microscopic model {$\uS$} is
isotropic. {$ZT$} is therefore also anisotropic.
Let now {$\underline{P}$}
be any symmetric and positive matrix and {$\lambda\ge 0$} a positive number; 
then by the properties of positive matrices, 
{$\underline{P}^{-1}\ge (\underline{P}+\lambda \underline{1})^{-1}$}. 
For the application to thermoelectrics let
\begin{equation}
\underline{\kappa}_e =
 \pmatrix{ \kappa_\parallel & \underline{\kappa}_{od}^T \cr
 \underline{\kappa}_{od} & \underline{\kappa}_\perp \cr}.
\label{93b}
\end{equation}
In analogy with Eq. (\ref{21a}) we find
\begin{equation}
\kappa_e^*=\kappa_\parallel - {\underline{\kappa}_{od}}^T
(\underline{\kappa}_\perp + \kappa_\ell \underline{1})^{-1} 
\underline{\kappa}_{od}.
\label{93c}
\end{equation}
Taking {$\underline{P}=\underline{\kappa}_\perp$} and 
{$\lambda = \kappa_\ell$}, the relation {$\underline{P}^{-1}\ge 
(\underline{P}+\lambda \underline{1})^{-1}$} shows that
\begin{equation}
{\underline{\kappa}_{od}}^T(\underline{\kappa}_\perp + \kappa_\ell
\underline{1})^{-1} \underline{\kappa}_{od} \le
{\underline{\kappa}_{od}}^T {\underline{\kappa}_\perp}^{-1} 
\underline{\kappa}_{od}.
\label{93d}
\end{equation}
Thus, for any {$\kappa_\ell$} 
\begin{equation}
\kappa_e^*(\kappa_\ell) \ge
\kappa_e^*(\kappa_\ell=0) = (\kappa_e/\sigma_0)\ceff.
\label{93e}
\end{equation}
This shows that the ratio {$\kappa_e^*/\ceff$} is
minimized when the axis of current flow in steady state lies 
along one of the principal directions, where, 
whatever the value of {$\kappa_\ell$}, the induced-field related terms 
vanish and so equality obtains in Eq. (\ref{93e}).
Writing {$ZT$} in the form
\begin{equation}
ZT = {S_0^2 \over {\kappa_e^*/T\ceff + \kappa_\ell/T\ceff}},
\label{96}
\end{equation}
the second term in the denominator is seen to be smallest along the
principal direction with largest electrical conductivity and therefore
$ZT$ is maximized for current flow along this direction.

On the other hand, for a sufficiently anisotropic lattice thermal 
conductivity the favored direction might be determined by its minimum
rather than that of the highest electrical conductivity.
For crystals in which two of the principal values of the electrical
conductivity are equal, such as $\rm Bi_2Te_3$ and SLs,
numerical results show that as long as the anisotropy
of {$\underline{\kappa}_\ell$} is smaller than that of {$\underline{\sigma}$},
the optimum {$ZT$} is still to be found in the
direction of greatest electrical conductivity.
This suggests that, generally speaking, the figure of merit will be
maximized in the principal crystal direction in which the ratio
$\sigma_i/\kappa_{\ell,i}$ is greatest, where $\sigma_i$ and
$\kappa_{\ell,i}$ are the principal values of the electrical and lattice
thermal conductivity tensors obtained from summing over valleys.

\section{Applications}
\label{sec:applications}

In the Section, the formalism of Sec.~\ref{sec:micro} is applied to
several systems in which the effective-mass approximation is valid
and for which, as shown above, the thermopower is isotropic.

\subsection{Bulk $\rm Bi_2Te_3$
\label{sec:BiTeBulk}}

Bismuth telluride is a semiconductor with a room-temperature energy gap
of 0.13~eV that crystallizes in a trigonal structure with space
group $\rm D^5_{3d}$ (R$\overline{3} m$).\cite{Madelung}
An orthogonal coordinate system can be defined in this structure,
consisting of the trigonal (three-fold rotation) axis,
a bisectrix axis that resides within a reflection plane and is normal
to the trigonal axis, and a binary axis which is along the two-fold
rotation axis orthogonal to the other two directions.
This coordinate system will be used throughout the paper and will be
referred to as the crystal frame.

As shown in Fig.~\ref{fig:BS}, the constant-energy surfaces of the
conduction band at low doping are ellipsoidal.
There are six degenerate valleys, each described by a highly
anisotropic effective-mass tensor.\cite{Goldsmid}
As illustrated in the figure, two of these valleys are bisected by the
trigonal-bisectrix plane and their principal axes are rotated
approximately 1.2$^\circ$ from the bisectrix axis.
The light effective masses are $m_1 = 0.025m_0$ near the bisectrix,
$m_2 = 0.26m_0$ along the binary axis and $m_3 = 0.19m_0$ near the
trigonal axis.
These two surfaces are related to each other by inversion through the origin.
The remaining four constant-energy surfaces are obtained from the first
two by rotations through $\pm 2 \pi / 3$ about the trigonal axis.

The tensorial transport distribution function, Eq.~(\ref{41}), 
involves the scalar part ${\cal T} (\eps)$ given by Eq.~(\ref{40}).
The relaxation time $\tau (\eps)$ taken to be independent of energy and
is determined from the experimental mobility along the bisectrix,
1200~cm$^2$~/~V~s.\cite{Goldsmid}
The matrix part $\uA$ is constructed from the experimentally derived
inverse mass matrix for a single conduction band ellipsoid
($n=1$).\cite{Goldsmid}.
Application of the point group operations of the crystal give the other
inverse mass matrices needed to construct $\uA$ in the crystal frame
[cf. Eq.~(\ref{39})], which is then transformed into the frame of the sample.
$ZT$ is obtained from the resulting transport coefficients by taking the
lattice thermal conductivity at 300~K to be 1.5~W~/~m~K
(Ref.~\onlinecite{Goldsmid}) and isotropic.
We shall assume a carrier density $n = 5.2 \times 10^{18}~{\rm cm}^{-3}$,
which gives the maximal $ZT = 0.71$ in our model when the sample is cut
along the bisectrix.

The magnitude of the induced fields in a crystal cut at an angle $\theta$
with respect to the trigonal axis in the trigonal-bisectrix plane is plotted
in Fig.~\ref{fig:BiTeInduced}(a).
The induced field vanishes when the sample is
cut along either the bisectrix or the trigonal directions.
This field reaches its maximum of 76\% of the external field for a sample
cut 0.36$\pi$ radians with respect to the trigonal axis.
As discussed in Sec.~\ref{sec:formalism}, an external temperature gradient may
induce temperature gradients along the transverse faces.
The induced gradients are at most 12\% of the external gradient and
have little effect on $ZT$.

The effective electrical conductivity $\sigma_{\rm eff}$ [Eq.~(\ref{24a})]
is shown in Fig.~\ref{fig:BiTeInduced}(b).
As the direction in which the sample is cut is changed from the
bisectrix ($\theta = \pi/2$) to the trigonal ($\theta = 0$),
the conductivities computed with
($\sigma_{\rm eff}$) and without ($\sigma_{xx}$) sample boundary effects
are both reduced by a factor of four due to the intrinsic anisotropy of
the material.
$\sigma_{\rm eff}$ is further reduced with respect to $\sigma_{xx}$ by
the induced fields when the sample is oriented along low-symmetry
directions by as much as 60\%.
The influence of the sample boundaries is therefore substantial.

The combination of intrinsic anisotropy and the effects of the induced
fields also affect $ZT$, but only if $\kappa_\ell$ is non-zero.
As shown in Fig.~\ref{fig:BiTeZT}(a) for a hypothetical material with
$\kappa_\ell = 0$, $ZT = 2.6$ and is isotropic [Eq.~(\ref{55})].
As seen in Fig.~\ref{fig:BiTeZT}(b), when $\kappa_\ell$ assumes its bulk
value, $ZT$ becomes anisotropic and decreases to a maximum value of 0.71.
The maximal $ZT$ applies to samples cut along the high-conductivity
bisectrix-binary plane, consistent with the results of Sec.~\ref{sec:bound}.
The minimal $ZT = 0.22$ occurs for samples cut along the low-conductivity
trigonal axis.
Despite the complicated many-valley band structure, $ZT$ is independent
of the azimuthal angle $\phi$ defined in the inset to
Fig.~\ref{fig:BiTeZT}(a).
This is consistent with a group-theoretical analysis and is suggested
by the relative orientations of the ellipsoids in Fig.~\ref{fig:BS}.

Introducing anisotropy into $\kappa_\ell$ by increasing its value along
the bisectrix reduces $ZT$ for samples cut along this direction, but not
for those cut along the trigonal axis.
When the ratio $\sigma / \kappa_\ell$ along the trigonal axis
exceeds that along the bisectrix, the maximal $ZT$ shifts to
the low-conductivity trigonal direction.
$ZT$ is never maximal along a low-symmetry direction.
These numerical results support the conjecture at the end of
Sec.~\ref{sec:bound}.

\subsection{HgTe/HgCdTe Superlattices
\label{sec:sl}}

We now consider HgTe/HgCdTe SLs, one of many SLs whose well and barrier
materials have direct $\Gamma$-point band gaps.
The constant energy surfaces of the lowest conduction subband $C1$
consist of single ellipsoids of revolution aligned with the growth
axis of the SL.
The conductivity will typically be a minimum along the growth axis
($\sigma_{\rm min}$) and a maximum within the SL plane ($\sigma_{\rm max}$).
If the sample is cut at an angle $\phi$ relative to the
SL planes, the induced fields will be non-zero only for the
faces intersected by the SL growth axis, as indicated in the inset
to Fig.~\ref{fig:slInduced}.
Use of Eq.~(\ref{13}) then yields the relative magnitude of the induced
electric field as
\begin{equation}
\frac{{\cal E}^{\rm ind}}{F^{\rm ext}} =
  \frac{(\sigma_{\rm max}-\sigma_{\rm min}) \sin \phi \cos \phi}
       {\sigma_{\rm max} \sin^2 \phi + \sigma_{\rm min} \cos^2 \phi} .
\label{eq:slInduced}
\end{equation}

This relation is plotted in Fig.~\ref{fig:slInduced} for several
values of the ratio $\sigma_{\rm min} / \sigma_{\rm max}$.
When the system is isotropic, $\sigma_{\rm min} / \sigma_{\rm max} = 1$ and
the induced field vanishes.
With increasing anisotropy, ${\cal E}^{\rm ind}$ becomes non-zero away
from the high-symmetry growth and in-plane orientations and peaks at an
angle
\begin{equation}
\phi_{\rm max} =
  \sin^{-1} \sqrt{\sigma_{\rm min}/(\sigma_{\rm max}+\sigma_{\rm min})}.
\label{eq:phimax}
\end{equation}
The maximum induced field is
\begin{equation}
\left. {\cal E}^{\rm ind}/F^{\rm ext} \right|_{\rm max} =
  (\sigma_{\rm max}-\sigma_{\rm min})/
  2 \sqrt{\sigma_{\rm max} \sigma_{\rm min}}.
\label{eq:indmax}
\end{equation}
In the limit $\sigma_{\rm min}/\sigma_{\rm max} << 1$, these
relations reduce to
$\phi_{\rm max} \sim \sqrt{\sigma_{\rm min}/\sigma_{\rm max}}$ and
${\cal E}^{\rm ind}/F^{\rm ext} |_{\rm max} \sim
(1/2) \sqrt{\sigma_{\rm max}/\sigma_{\rm min}}$.
Thus, as illustrated in Fig.~\ref{fig:slInduced}, a relative induced field
much larger than unity is obtained for sufficiently small
$\sigma_{\rm min}/\sigma_{\rm max}$.
From Eq.~(\ref{eq:indmax}), the maximal induced field exceeds the
external one when
$\sigma_{\rm min}/\sigma_{\rm max} \le 3 - 2\sqrt{2} \cong 0.172$.
This geometrical effect may be practically useful.

As specific examples, consider two superlattices:
(1) 30-\AA~HgTe~/ 20-\AA~$\rm Hg_{0.15}Cd_{0.85}Te$ and
(2) 20-\AA~HgTe~/ 40-\AA~$\rm Hg_{0.15}Cd_{0.85}Te$.
Realistic band structures for these materials are obtained
through an 8-band $\bf K \cdot p$ theory within the
envelope function approximation\cite{Ehrenreich} based on the parameters
in Tab.~\ref{tab:slParameters}.
Application of this approach to other II-VI and III-V SLs 
yields accurate band structures which reproduce experimental optical
absorption spectra with no adjustable parameters.\cite{Ehrenreich}
The masses of the $C1$ subband along the growth and
in-plane directions required for the transport calculations
are given in Tab.~\ref{tab:slResults}.
The room-temperature energy gaps are seen to satisfy
$E_g \ge 10 k_B T$ in both structures, so hole conduction is negligible.
Transport in the next highest conduction subband is also negligible, since
it lies more than $20 k_B T$ higher in energy than $C1$ in both SLs.
The relaxation time and lattice thermal conductivity enter $ZT$ significantly
only in the ratio $\kappa_\ell/ \sigma_{\rm eff} T$ [Eq.~(\ref{96})].
Their values are estimated from those of an alloy of the same composition
as the SL (Tab.~\ref{tab:slResults}).
Thus, interface scattering is neglected.

The thermoelectric performance of these SLs is shown in
Tab.~\ref{tab:slResults}.
Since $\sigma_{\rm min} / \sigma_{\rm max} = m_\| / m_\bot$, SL (1)'s 0.66
mass ratio between the in-plane and growth directions yields a maximum
induced field of only $0.2 F^{\rm ext}$.
SL (2)'s much smaller 0.39 mass ratio, indicating a larger anisotropy,
yields a correspondingly larger induced field of $0.5 F^{\rm ext}$.
For even larger anisotropies, the effective mass approximation breaks
down, but estimates using the band structure of Sec.~\ref{sec:anisS}
suggest that induced fields up to $17 F^{\rm ext}$
can be obtained in a 40-\AA~HgTe / 50-\AA~CdTe SL.
As for bulk $\rm Bi_2Te_3$, induced temperature gradients are small, at most
2\% of the external gradient for SL~(1) and SL~(2).
Note that the maximal $ZT$s shown in Tab.~\ref{tab:slResults}, although
small in magnitude, represent a substantial increase over the equivalent
bulk alloy.

\subsection{$\rm Bi_2Te_3$ Quantum Wells
\label{sec:BiTeQW}}

Consider a quantum well formed from a multi-valley semiconductor.
Within the single-band envelope function approximation,\cite{Bastard}
the wave function for the $n$th valley, $\psi^{(n)} ({\bf r})$,
satisfies
\begin{equation}
\left[ -\sum_{ij} \alpha_{ij}^{(n)} \partial_i \partial_j + V({\bf r}) \right]
\psi^{(n)} ({\bf r}) = E^{(n)} \psi^{(n)}({\bf r}),
\label{eq:SE}
\end{equation}
where $\alpha_{ij} = (\hbar^2/2) M_{ij}^{(n)-1}$ is proportional to the bulk
inverse effective mass tensor $M_{ij}^{(n)-1}$,
$E^{(n)}$ the energy measured from
the bottom of the valley in the bulk, and $i,j = x$, $y$, or $z$.
$V({\bf r})$ is the confining potential, which is taken to be a square
well of width $L_z$ and infinite height.
Solution of Eq.~(\ref{eq:SE}) yields the eigenfunctions
\begin{equation}
\psi_m^{(n)} ({\bf r}) =
  \sqrt{2/L_z} \exp \left[ i \left( k_x x + k_y y
  - (\alpha_{zx}^{(n)} k_x + \alpha_{zy}^{(n)} k_y) z / \alpha_{zz}^{(n)}
  \right) \right]
  \sin (m \pi z/L_z)
\label{eq:wf2D}
\end{equation}
and the eigenenergies
\begin{equation}
E_m^{(n)} (k_x,k_y) =
  \sum_{i,j = x,y} k_i \alpha_{ij}^{(n){\rm Well}} k_j
  + E_m^{(n){\rm Conf}}
\label{eq:e2D}
\end{equation}
with 
\begin{equation}
\alpha_{ij}^{(n){\rm Well}} = 
\left( \begin{array}{cc}
  \alpha_{xx}^{(n)}-\alpha_{xz}^{(n)}\alpha_{zx}^{(n)}/\alpha_{zz}^{(n)} &
  \alpha_{xy}^{(n)}-\alpha_{xz}^{(n)}\alpha_{zy}^{(n)}/\alpha_{zz}^{(n)} \\
  \alpha_{yx}^{(n)}-\alpha_{yz}^{(n)}\alpha_{zx}^{(n)}/\alpha_{zz}^{(n)} &
  \alpha_{yy}^{(n)}-\alpha_{yz}^{(n)}\alpha_{zy}^{(n)}/\alpha_{zz}^{(n)} \\
\end{array} \right)
\label{eq:mass2D}
\end{equation}
and the confinement energy
\begin{equation}
E_m^{(n){\rm Conf}} = \alpha_{zz}^{(n)} \pi^2 m^2 / L_z^2,
\label{eq:Econf2D}
\end{equation}
for $m = 1,2,3,...$

Thus, the quantum well band structure differs from that of the bulk
in two ways.
First, the energy of the bottom of the $n$th valley is shifted with respect
to the bulk by an amount $E_1^{(n)\rm Conf}$ that is inversely proportional
to $m_{zz}^{(n)}$.
Since $m_{zz}^{(n)}$ is generally different for different $n$, the
confinement splits the valley degeneracy.
If the well is sufficiently narrow, only the valleys having the lowest
confinement energy will contribute to the transport.
Second, as seen from Eq.~(\ref{eq:mass2D}), the masses in the quantum
well differ from those in the bulk by terms second order in the off-diagonal
elements of the bulk inverse effective-mass tensor.

Suppose the well material is n-type $\rm Bi_2Te_3$.
The bulk band structure was described in Sec.~\ref{sec:BiTeBulk}.
The relaxation time and lattice thermal conductivity can be obtained
as in Ref.~\onlinecite{Hicks2D}, which neglects interface effects.
Eqs.~(\ref{eq:mass2D}), (\ref{41}), and (\ref{72})-(\ref{73}) then lead
directly to the transport coefficients and $ZT$.
For simplicity, $\kappa_\ell$ of the barriers is neglected.
Thus, $ZT$ is an overestimate.\cite{MahanQW,QWnote}

Examine two orientations of the quantum well:
one in which the growth direction is along the trigonal axis and one
in which it is along the binary.
In both cases, take the currents to flow along the bisectrix.
These choices are expected to yield the best $ZT$s, since the currents
are along the bulk high-conductivity direction and the alignment of the
well with the crystal axes eliminates detrimental induced fields.
The maximal $ZT$ for these configurations is plotted as a function of
well width in Fig.~\ref{fig:zt2D}.

As seen in this figure, quantum well growth along the trigonal axis yields
larger $ZT$s.
From Fig.~\ref{fig:BS}, the masses of each ellipsoid along the
trigonal axis are the same, but they differ along the binary axis.
Thus, all six ellipsoids contribute to transport in wells grown along the
trigonal axis, but valley splitting allows only two ellipsoids to contribute
in wells grown along the binary axis.
The resulting larger density of states in the former case leads to
higher $ZT$s.
For a given orientation, the density of states is also larger for smaller
well widths, resulting in an increase of $ZT$ with decreasing well widths
in both orientations.\cite{Hicks2D}

For comparison, Fig.~\ref{fig:zt2D} also shows $ZT$ computed using
the simpler $\rm Bi_2Te_3$ band structure of Ref.~\onlinecite{Hicks2D}.
This band structure consists of six equivalent ellipsoids whose principal
axes are all aligned with those of the crystal.
As in the multi-ellipsoid model, $ZT$ is larger for wells grown along the
trigonal axis and for smaller well widths.
However, the magnitude of $ZT$ is larger than in the multi-ellipsoid model.
This difference arises because each ellipsoid shares the same
orientation, allowing all of the ellipsoids' low-mass directions to lie
along the bisectrix and preventing any valley splitting.

\section{Microscopic Model:
Non-Parabolic Bands}
\label{sec:anisS}

Within the effective-mass approximation, the analysis of Sec.~\ref{sec:micro}
showed that the thermopower and Lorentz number are isotropic.
To see how a non-parabolic band structure affects these conclusions,
consider an electronic dispersion relation of the Esaki-Tsu form:\cite{Esaki}
\begin{equation}
\varepsilon({\bf k}) = \hbar^2 k_\|^2/2 m_\| + \Delta (1 - \cos k_z d)
\label{eq:Esaki}
\end{equation}
with wave vector ${\bf k} = (k_\| \cos \varphi, k_\| \sin \varphi, k_z)$.
This relation models a superlattice of period $d$.
The in-plane dispersion is parabolic with mass $m_\|$, and that
along the growth direction has a tight-binding form with band width
$2 \Delta = (\varepsilon(0,0,\pi/d)-\varepsilon(0,0,0))$.
Since the mass along the growth direction $m_z = \hbar^2 / \Delta d^2$,
the anisotropy can be increased by reducing $\Delta$.

In the principal frame of the SL, the transport distribution tensor
[Eq.~(\ref{27})] is diagonal with components 
\begin{equation}
\Sigma_{\rm Growth} (\varepsilon) =
  \frac{m_\| d \tau(\varepsilon)}{2 \pi^2 \hbar^4}
  \left\{ \begin{array}{cc}
  \Delta^2 \cos^{-1} (1-\varepsilon/\Delta)
    + (\varepsilon-\Delta) \sqrt{\varepsilon (2 \Delta - \varepsilon)}, &
  \varepsilon < 2 \Delta \\
  \pi \Delta^2, & \varepsilon \ge 2 \Delta \end{array} \right.
\label{eq:TDFgrowth}
\end{equation}
along the growth direction and
\begin{equation}
\Sigma_{\rm In-plane} (\varepsilon) =
  \frac{\tau(\varepsilon)}{\pi^2 \hbar^2 d}
  \left\{ \begin{array}{cc}
  (\varepsilon-\Delta) \cos^{-1} (1-\varepsilon/\Delta)
    + \sqrt{\varepsilon (2 \Delta - \varepsilon)}, &
  \varepsilon < 2 \Delta \\
  \pi (\varepsilon-\Delta), & \varepsilon \ge 2 \Delta \end{array} \right.
\label{eq:TDFinplane}
\end{equation}
along the planes.
The transport coefficients are obtained from Eqs.~(\ref{26prime})-(\ref{26})
in the principal frame and then transformed into the sample frame.
As in Sec.~\ref{sec:micro}, the relaxation time $\tau ({\bf k})$ is assumed
to be a function of energy only.
Direct computation with $\tau \propto \varepsilon^r$ indicates that
the qualitative features discussed below are independent of the choice of $r$.
Quantitatively, the thermopower is an approximately linear function of $r$
at fixed sample orientation and chemical potential $\mu$ and increases by
approximately 50\% as $r$ goes from 0 to 1.5.
In what follows, $r = 0$, $d = $100~\AA, $m_\|/m_0 = 0.021$,
and $\Delta = 57$~meV, corresponding to the C1 subband in a
50-\AA~$\rm Hg_{0.75}Cd_{0.25}Te$ / 50-\AA~$\rm Hg_{0.7}Cd_{0.3}Te$ SL.

The anisotropy in the resulting effective thermopower $S_{\rm eff}$
[Eq.~(\ref{21})] and Lorentz number
$L_{\rm 0,eff} = \kappa_{\rm eff} / \sigma_{\rm eff} T$
is shown in Fig.~\ref{fig:anisS}.
For $\mu < 0$, $S_{\rm eff}$ along the growth and in-plane directions
differ by $< 10$\%.
This near isotropy is expected, since the carriers determining $S_{\rm eff}$
are near the zone center, where the effective-mass approximation is good.
The anisotropy increases substantially as $\mu$ increases past $2 \Delta$,
reaching over 6000 at $\mu = 0.4$~eV.
From Fig.~\ref{fig:anisS}(b), the anisotropy in $L_{\rm 0,eff}$ is at most
30\% and goes to zero in the large-$\mu$ limit, approaching the metallic
value of $(\pi^2/3) (k_B/e)^2$.
As $ZT$ for this band structure is maximal for $\mu \sim 0$, the anisotropy
in $S_{\rm eff}$ and $L_{\rm 0,eff}$ in the relevant parameter range
are small.

These numerical results are also consistent with the idea that the
induced fields are detrimental to $ZT$, as discussed in Appendix~A.
As shown by the dotted lines in Fig.~\ref{fig:anisS}(a), the induced fields
present along low-symmetry directions reduce $S_{\rm eff}$ below $S_{xx}$.
However, $S_{\rm eff}$ is always bounded by $S_{\rm eff} = S_{xx}$ along
the principal directions.
Thus, optimal thermoelectric performance is achieved for samples cut along
the principal axes.

\section{Summary}
\label{sec:conclusions}

This paper developed the transport theory relevant for anisotropic,
multi-valleyed materials, taking into account the full tensorial
structure of the electronic transport coefficients and including the
effects of sample boundaries.
Induced transverse fields are associated with samples cut along directions
in which the conductivity had off-diagonal elements.
These fields can be larger than the applied fields.
Within the effective-mass and relaxation-time approximations and the
neglect of intervalley scattering, the tensor characterizing the structure
appears as a simple multiplicative factor of each of the transport
coefficients.
This factorization results in an isotropic thermopower and Lorentz number
even in the presence of anisotropy.
In a hypothetical material having no lattice thermal conductivity
$ZT$ would therefore be isotropic.
For non-vanishing and sufficiently isotropic lattice thermal conductivity
$ZT$ is maximal for samples cut along the high-conductivity directions.
Widely ranging numerical results suggest that the maximal $ZT$ generally
occurs along the principal direction which maximizes the ratio of the
electronic to lattice thermal conductivity.
An upper bound for $ZT$ is given which generalizes the
result of Mahan and Sofo\cite{Mahan} to anisotropic systems.

Application of these formal results to some systems of interest showed the
following.
(1)  Bulk n-type $\rm Bi_2Te_3$ exhibits induced transverse
fields, reduced effective conductivity, and anisotropic $ZT$s.
These effects should be easily observable.
(2)  In $\rm HgTe/Hg_{1-x}Cd_xTe$ SLs, whose anisotropy is adjustable
by varying the layer widths and composition, increased anisotropy is
associated with larger induced fields.
(3)  The confining potential in {\it isolated} $\rm Bi_2Te_3$ quantum
wells splits the valley degeneracy and changes the masses from their
bulk values.
The latter features suggest that optimal $ZT$s are associated with wells
grown along the trigonal direction.
(4)  The effect of non-parabolic dispersion on the thermopower is very
small for the doping range over which $ZT$ is maximal.
The maximal thermopower occurs along principal directions,
which supports the suggestion given in the previous paragraph.

\acknowledgements

Discussions with E. Runge are gratefully acknowledged.
This work was supported by DARPA through ONR Contract No.~N00014-96-1-0887
and the NSF through Che9610501.

\appendix
\section{}

In this appendix we present a unified form of the linear-response theory
which, because of its explicit use of the Onsager coefficients, has the
advantage of treating the electric and heat currents more 
symmetrically than in Sec.~\ref{sec:formalism}.
Note that, since the Onsager coefficients are assumed to be symmetric
matrices, this formalism does not apply to monoclinic, triclinic, or
chiral materials.
The generalized fluxes $\bf J$ are given in terms of the generalized forces
$\bf f$ by
\begin{equation}
{\bf J} = \underline{L} {\bf f}
\label{1000}
\end{equation}
with
\begin{equation}
{\bf J} = \left( \begin{array}{cccccc}
  J^e_x & J^Q_x & J^e_y & J^Q_y & J^e_z & J^Q_x \end{array} \right)^T =
\pmatrix{ {\bf J}_\parallel \cr {\bf J}_\perp \cr}
\end{equation}
and
\begin{equation}
{\bf f} =  \left( \begin{array}{cccccc}
  e{\cal E}_x & -\del_xT/T & e{\cal E}_y & -\del_yT/T &
  e{\cal E}_z & -\del_zT/T
  \end{array} \right)^T  =
\pmatrix{ {\bf f}_\parallel \cr {\bf f}_\perp \cr}
\end{equation}
where {${\bf J}_\parallel$} and {${\bf f}_\parallel$} contain the first
two components and {${\bf J}_\perp$} and {${\bf f}_\perp$} the remaining
four components of {${\bf J}$} and {${\bf f}$}, respectively. With respect
to these conventions the matrix of Onsager coefficients, {$\underline{L}$},
assumes the form
\begin{equation}
\underline{L} = \pmatrix{ \underline{L}_\parallel & \underline{L}_{od}^T \cr
\underline{L}_{od} & \underline{L}_\perp \cr}
\end{equation}
with
\begin{equation}
\underline{L}_\parallel = \pmatrix{ L^{11}_{xx} & L^{21}_{xx} \cr
                                    L^{12}_{xx} & L^{22}_{xx} \cr},
\end{equation}
\begin{equation}
\underline{L}_{od} = \left( \begin{array}{cccc}
 L^{11}_{xy} & L^{21}_{xy} & L^{11}_{xz} & L^{21}_{xz} \\
 L^{12}_{xy} & L^{22}_{xy} & L^{12}_{xz} & L^{22}_{xz}
\end{array} \right)^T
\end{equation}
and
\begin{equation}
\underline{L}_\perp = 
\pmatrix{ L^{11}_{yy} & L^{21}_{yy} & L^{11}_{yz} & L^{21}_{yz} \cr
          L^{12}_{yy} & L^{22}_{yy} & L^{12}_{yz} & L^{22}_{yz} \cr
          L^{11}_{zy} & L^{21}_{zy} & L^{11}_{zz} & L^{21}_{zz} \cr
          L^{12}_{zy} & L^{22}_{zy} & L^{12}_{zz} & L^{22}_{zz} \cr}.
\end{equation}
Equation (\ref{1000}) and the condition that {${\bf J}_\perp=0$} yield
\begin{equation}
{\bf f}_\perp = -\underline{L}_\perp^{-1} \underline{L}_{od} {\bf f}_\parallel;
\end{equation}
hence,
\begin{equation}
{\bf J}_\parallel = \underline{L}_{\rm eff} {\bf f}_\parallel
\end{equation}
with
\begin{equation}
\underline{L}_{\rm eff} = \underline{L}_\parallel - \underline{L}_{od}^T
\underline{L}_\perp^{-1} \underline{L}_{od}
\equiv \underline{L}_\parallel - \underline{N}_\parallel
\end{equation}
It can be shown that {$\underline{N}_\parallel>0$} and 
{$\underline{L}_{\rm eff}\ge 0$}.
We may decompose {${\bf J}_\parallel$} into external and induced parts
as follows:
\begin{equation}
{\bf J}_\parallel = {\bf J}^{\rm ext} + {\bf J}^{\rm ind}
\end{equation}
with
\begin{equation}
{\bf J}^{\rm ext} = \underline{L}_\parallel {\bf f}_\parallel
\end{equation}
and
\begin{equation}
{\bf J}^{\rm ind} = -\underline{N}_\parallel {\bf f}_\parallel.
\end{equation}
In Sec.~\ref{sec:formalism} it was found that, for an applied electric field
only ({${\bf f}_\parallel=(e{\cal E}_x, 0)$}), the induced electric current
opposes the external electric current because 
{$\sigma_{\rm eff} \le \sigma_\parallel$}, and likewise for an applied thermal
gradient only, the induced heat current opposes the external heat current.
Physical intuition suggests that these results should generalize to the
case where both an applied electric field and applied thermal gradient are
present.
Then we expect that, if both components of {\bf f} have the same sign,
{${\bf J}^{\rm ind}$} should oppose 
{${\bf J}^{\rm ext}$} in the sense that they lie in opposite quadrants
of the plane spanned by {$J^e_x$} and {$J^Q_x$}.
Although we have not proven this conjecture, it is consistent with numerical
results for randomly constructed matrices of Onsager coefficients.
Undoubtedly a proof requires a general microscopic model.
Finally, we note that the entropy generation is given by
\begin{equation}
\left. -T {{\partial S} \over {\partial t}}\right|_{\rm field}
= {\bf J} \cdot {\bf f} = {\bf J}^{\rm ext} \cdot {\bf f}_\parallel
+ {\bf J}^{\rm ind} \cdot {\bf f}_\parallel.
\end{equation}
Since {$\underline{L}_\parallel > 0$} and {$\underline{N}_\parallel > 0$},
we have {${\bf J}^{\rm ext} \cdot {\bf f}_\parallel > 0$} and
{${\bf J}^{\rm ind} \cdot {\bf f}_\parallel < 0$}. Thus, external currents
{\it decrease} {$\dot{S}|_{\rm field}$} but induced currents {\it increase}
{$\dot{S}|_{\rm field}$}.

\section{}

We present here a derivation of an upper bound of the Sofo and Mahan
form\cite{Mahan} for an anisotropic material.
From
{$\usigma = \uA \sigma_0$} we infer
\begin{equation} \ceff
=\sigma_0 a_0
\label{78}
\end{equation}
with
\begin{equation}
a_0 = A_{xx} -
A_{xy}\frac{A_{yx}A_{zz}-A_{yz}A_{zx}}{A_{yy}A_{zz}-A_{yz}A_{zy}}-
A_{xz}\frac{A_{zx}A_{yy}-A_{zy}A_{yx}}{A_{zz}A_{yy}-A_{yz}A_{zy}} .
\label{79}
\end{equation}
When the {$\kappa_\ell$}-dependence in Eq. (\ref{86})
is taken into account, we find similarly
\begin{equation}
\kappa_e^* = \kappa_e a(\kappa_\ell)
\label{87}
\end{equation}
with
{$a(\kappa_\ell=0)=a_0$} and
\begin{eqnarray}
a(\kappa_\ell) &=A_{xx} &-
A_{xy}{{A_{yx}(A_{zz}+\kappa_\ell/\kappa_e)-A_{yz}A_{zx}} \over
{(A_{yy}+\kappa_\ell/\kappa_e)(A_{zz}+\kappa_\ell/\kappa_e)-A_{yz}A_{zy}}}
\nonumber \\
&&-
A_{xz}{{A_{zx}(A_{yy}+\kappa_\ell/\kappa_e)-A_{zy}A_{yx}} \over
{(A_{yy}+\kappa_\ell/\kappa_e)(A_{zz}+\kappa_\ell/\kappa_e)-A_{yz}A_{zy}}}
.
\label{88}
\end{eqnarray}
Therefore
\begin{equation}
ZT = {{T\sigma_0S_0^2a_0} \over
{(\kappa_0-T\sigma_0S_0^2)a(\kappa_\ell)+\kappa_\ell}}.
\label{89}
\end{equation}
Following Mahan and Sofo, introduce dimensionless integrals
\begin{equation}
I_n = \int_{-\mu/k_B T}^\infty dx
{{e^x} \over {(e^x+1)^2}}s(x)x^n, \qquad s(x)= \hbar r_0
{\cal T}(\mu + x kT),
\label{82}
\end{equation}
where {$r_0$} is the Bohr radius. In terms
of these moments,
\begin{eqnarray}
\sigma_0 &=& \ts I_0 \cr
\sigma_0 S_0 &=& ( {k_B / e} ) \ts I_1 \\
\kappa_0 &=& ( {k_B / e} )^2 T \ts I_2
\label{83}
\end{eqnarray}
where {$\ts = e^2/\hbar r_0$} has dimensions of conductivity. Then
\begin{eqnarray}
ZT &=&
{{T(k_B/e)^2\ts^2I_1^2/\ts I_0} \over 
{((k_B/e)^2T\ts I_2 - T(k_B/e)^2 \ts I_1^2/I_0)a(\kappa_\ell)/a_0
+ \kappa_\ell/a_0}} \\
&=& {{\tilde{\alpha} I_1^2/I_0} \over 
{(\tilde{\alpha} I_2 - \tilde{\alpha} I_1^2/I_0)a(\kappa_\ell)/a_0 + 1/a_0}} \\
&=& {\xi \over {(1 - \xi)a(\kappa_\ell)/a_0 + B}}
\label{84}
\end{eqnarray}
with {$\tilde{\alpha} = (k_B/e)^2T\ts /\kappa_\ell$},
\begin{equation}
\xi = I_1^2/I_0I_2,
\label{eq:xi}
\end{equation}
and $B=1/\tilde{\alpha} I_2 a_0 = \kappa_\ell/\kappa_0 a_0$.
By the Cauchy-Schwarz
inequality, {$0 \le \xi \le 1$}. The limit as {$\xi$} tends to 1 maximizes
the figure of merit by maximizing the numerator and minimizing the
denominator in Eq. (\ref{84}).
From Eq. (\ref{88}) it may be seen that for
{$\kappa_\ell/\kappa_e = \kappa_\ell/(k_B/e)^2T\ts I_2(1-\xi)$} sufficiently
large, {$a(\kappa_\ell)$} tends to {$A_{xx}$}, and certainly this
condition is met as {$1-\xi$} tends to zero. Thus as {$\xi \rightarrow 1$}
we have that
\begin{equation}
ZT \rightarrow {\xi \over {(1-\xi)A_{xx}/a_0+B}}
\le {1 \over B} = {{\kappa_0 a_0} \over {\kappa_\ell}}.
\label{91}
\end{equation}


\newpage
\begin{table}
\caption{
Room temperature energy gap $E_g$, spin-orbit splitting $\Delta$,
and heavy-hole effective mass $m_{HH}$ for a
$\rm HgTe/Hg_{0.15}Cd_{0.85}Te$ superlattice.
The valence band offset is 0.30~eV from a linear interpolation of the
value in Ref.~\protect\onlinecite{Johnson}, and
the momentum matrix element $(2/m_0) |<Z|p_z|S>|^2$ = 18.0~eV
(Ref.~\protect\onlinecite{Dornhaus}).
\label{tab:slParameters}}
\begin{center}
\begin{tabular}{lccc}
 & HgTe & $\rm Hg_{0.15}Cd_{0.85}Te$ \\
\hline
$E_g$ at 300~K (eV)$^a$ & -0.166 & 1.191 \\
$\Delta$ (eV)$^b$ & 1.0 & 0.9  \\
$m_{HH}/m_0^{c}$ & 0.7 & 0.7  \\
\end{tabular}
\end{center}
$^a$Ref.~\protect\onlinecite{Brice}.

$^b$Ref.~\protect\onlinecite{Kane}.

$^c$Ref.~\protect\onlinecite{Schulman}.

\end{table}

\begin{table}
\caption{
Parameters and thermoelectric performance in HgCd/HgCdTe superlattices (SLs).
$m_\bot$ and $m_\|$ are the $C1$ effective masses along and normal to the
growth axis, $({\cal E}^{\rm ind}/F^{\rm ext})_{\rm max}$ is the maximal
relative magnitude of the induced field, $\phi_{\rm max}$ is the angle
relative to the SL planes at which the sample must be cut to obtain this
field, $(ZT)_{\rm max}$ is the maximal figure of merit for the SL,
and $(ZT)_{\rm alloy}$ is that for the equivalent alloy.
\label{tab:slResults}}
\begin{center}
\begin{tabular}{lcc}
 & SL (1) & SL (2) \\
 & 30-\AA~HgTe / 
 & 20-\AA~HgTe / \\
 & 20-\AA~$\rm Hg_{0.15}Cd_{0.85}Te$
 & 40-\AA~$\rm Hg_{0.15}Cd_{0.85}Te$ \\
\hline
Band structure:$^a$\\
\makebox[0.5in]{ }
$E_g$ (eV) & 0.25 & 0.46 \\
\makebox[0.5in]{ }
$m_\bot / m_0$ & 0.038 & 0.110 \\
\makebox[0.5in]{ }
$m_\| / m_0$ & 0.025 & 0.043 \\
\makebox[0.5in]{ }
$m_\| / m_\bot$ & 0.66 & 0.39 \\
Transport parameters:\\
\makebox[0.5in]{ }
$\tau$~(s)$^b$ & $5.3 \times 10^{-14}$ & $2.6 \times 10^{-14}$ \\
\makebox[0.5in]{ }
$\kappa_\ell$~(mW / cm K)$^c$ & 8.2 & 12 \\
Thermoelectric performance:$^d$\\
\makebox[0.5in]{ }
$({\cal E}^{\rm ind} / F^{\rm ext})_{\rm max}$ & 0.21 & 0.48 \\
\makebox[0.5in]{ }
$\phi_{\rm max}$ (rad) & 0.22$\pi$  & 0.18$\pi$ \\
\makebox[0.5in]{ }
$(ZT)_{\rm max}$ & 0.14 & 0.086 \\
\makebox[0.5in]{ }
$(ZT)_{\rm alloy}$ & 0.12 & 0.061 \\
\end{tabular}
\end{center}
$^a$Based on the parameters in Tab.~\protect\ref{tab:slParameters}.

$^b$From the mobility data presented in Ref.~\protect\onlinecite{Pelletier}.

$^c$Quadratic fit to the data in Ref.~{\protect\onlinecite{kappaL}}.

$^d$This work.
\end{table}

\newpage
\begin{figure}
\caption{
Constant-energy surfaces of the conduction band valleys in $\rm Bi_2Te_3$.
The bisectrix, binary, and trigonal directions are indicated.
\label{fig:BS}}
\end{figure}

\begin{figure}
\caption{
Effect of induced transverse fields in bulk n-type $\rm Bi_2Te_3$.
(a) Normalized induced field $F_z/F_x$ [Eq.~(\protect\ref{13})]
and (b) microscopic ($\sigma_{xx}$, dot-dashed line) and effective
($\sigma_{\rm eff}$, solid line) conductivities [Eq.~(\protect\ref{24a})]
for the sample geometry shown in the inset.
\label{fig:BiTeInduced}}
\end{figure}

\begin{figure}
\caption{
$ZT$ of n-type $\rm Bi_2Te_3$ at 300~K for (a) a hypothetical
material with $\kappa_\ell = 0$ and (b) $\kappa_\ell$ = 1.5~W~/~m~K,
the bulk in-plane value, as a function of the direction in which the
sample is cut (defined in the inset).
\label{fig:BiTeZT}}
\end{figure}

\begin{figure}
\caption{
Relative magnitude of the induced electric field
${\cal E}^{\rm ind}/F^{\rm ext}$ in a superlattice as
a function of the angle $\phi$ defined in the inset for several values
of the conductivity ratio $\sigma_{\rm min}/\sigma_{\rm max}$
[Eq.~(\protect\ref{eq:slInduced}].
\label{fig:slInduced}}
\end{figure}

\begin{figure}
\caption{
Maximal $ZT$ at 300~K for an isolated $\rm Bi_2Te_3$ quantum well
within the multi-ellipsoid model of Sec.~\protect\ref{sec:BiTeQW}
as a function of well width for wells grown along the trigonal
(heavy dot-dashed line) or binary (heavy solid line) directions.
Results for the equivalent-ellipsoid model of
Ref.~\protect\onlinecite{Hicks2D} for wells grown along the trigonal
(thin dot-dashed line) and binary (thin solid line) directions are
also shown.
\label{fig:zt2D}}
\end{figure}

\begin{figure}
\caption{
(a) Effective thermopower $S_{\rm eff}$ and (b) effective Lorentz number
$L_{\rm 0,eff} = \kappa_{\rm eff} / \sigma_{\rm eff} T$ as a function of
chemical potential for the superlattice described in
Sec.~\protect\ref{sec:anisS}.
Shown are samples cut along $\phi = 0$ (heavy solid line), $\pi/4$
(heavy dotted line), and $\pi/2$ (heavy dashed line) as defined in
the inset to Fig.~\protect\ref{fig:slInduced}.
The results for $\phi = \pi/4$ neglecting the effects of the induced
fields are also presented (light dotted line).
The metallic limit of $L_0 / (k_B/e)^2 = \pi^2/3$ is indicated in (b).
\label{fig:anisS}}
\end{figure}


\begin{references}

\bibitem{Hicks2D}
L. D. Hicks and M. S. Dresselhaus, Phys. Rev. B {\bf 47}, 12727 (1993).

\bibitem{MahanQW}
G. D. Mahan and H. B. Lyon, Jr., J. Appl. Phys. {\bf 76}, 1899 (1994).

\bibitem{SofoSL}
J. O. Sofo and G. D. Mahan, Appl. Phys. Lett. {\bf 65}, 2690 (1994).

\bibitem{Broido95}
D. A. Broido and T. L. Reinecke, Phys. Rev. B {\bf 51}, 13797 (1995).

\bibitem{Hicks96}
L. D. Hicks, T. C. Harman, X. Sun, and M. S. Dresselhaus,
Phys. Rev. B {\bf 53}, R10493 (1996).

\bibitem{Broido97}
D. A. Broido and T. L. Reinecke, Appl. Phys. Lett. {\bf 70}, 2834 (1997)
and unpublished.

\bibitem{HicksBi}
L. D. Hicks, T. C. Harman, and M. S. Dresselhaus,
Appl. Phys. Lett. {\bf 63}, 3230 (1993).

\bibitem{BroidoBi}
D. A. Broido and T. L. Reinecke, Appl. Phys. Lett. {\bf 67}, 1170 (1995).

\bibitem{Hicks1D}
L. D. Hicks and M. S. Dresselhaus, Phys. Rev. B {\bf 47}, 16631 (1993).

\bibitem{Broido1D}
D. A. Broido and T. L. Reinecke, Appl. Phys. Lett. {\bf 67}, 100 (1995).

\bibitem{Mahan}
G. D. Mahan and J. O. Sofo,
Proc. Natl. Acad. Sci. USA {\bf 93}, 7436 (1996).

\bibitem{Madelung}
{\it Semiconductors:  Other than Group IV Elements and III-V Compounds},
edited by O. Madelung (Springer, New York, 1992).

\bibitem{Goldsmid}
H. J. Goldsmid, {\it Thermoelectric Refrigeration} (Plenum, New York, 1964).

\bibitem{Ehrenreich}
N. F. Johnson, H. Ehrenreich, P. M. Hui, and P. M. Young,
Phys. Rev. B {\bf 41}, 3655 (1990);
M. E. Flatt\'{e}, C. H. Grein, H. Ehrenreich, R. H. Miles, and H. Cruz,
J. Appl. Phys. {\bf 78}, 4552 (1995)
and references therein.

\bibitem{Bastard}
G. Bastard, in {\it Proceedings of the NATO Advanced Study Institute
on Molecular Beam Epitaxy in Heterostructures, Erice, July, 1983},
edited by L. L. Chang and K. Ploog (Martinus-Nijhoff, Dordrecht, 1984),
p. 381.

\bibitem{QWnote}
Actual quantum wells are embedded in another solid whose finite
thermal conductivity acts as a parasitic effect that lowers $ZT$.
Initially optimistic $ZT$ estimates in these systems\protect\cite{Hicks2D}
have been revised downward after including this
effect.\protect\cite{MahanQW,SofoSL}

\bibitem{Esaki}L. Esaki and R. Tsu, IBM J. Res. Dev. {\bf 14}, 61 (1970).

\bibitem{Brice}
J. C. Brice, in {\it Properties of Mercury Cadmium Telluride},
edited by J. Brice and P. Capper, EMIS Datareview Series No. 3
(INSPEC, New York, 1987), Ch. 5.1.

\bibitem{Johnson}
N. F. Johnson, P. M. Hui, and H. Ehrenreich,
Phys. Rev. Lett. {\bf 61}, 1993 (1988);
P. M. Hui, N. F. Johnson, and H. Ehrenreich,
J. Vac. Sci. Technol. A {\bf 7}, 424 (1989).

\bibitem{Dornhaus}
R. Dornhaus and D. Nimtz, in {\it Narrow-Gap Semiconductors},
edited by G. H\"{o}hler (Springer, New York, 1983).

\bibitem{Kane}
E. O. Kane, in {\it Narrow Gap Semiconductors:  Physics and Applications},
edited by W. Zawadzki, Lecture Notes in Physics Vol. 133
(Springer, New York, 1981), p. 19.

\bibitem{Schulman}
J. N. Schulman and Y.-C. Chang, Phys. Rev. B {\bf 33}, 2594 (1986).

\bibitem{Pelletier}
R. Granger and C. M. Pelletier,
J. Crys. Growth {\bf 138}, 486 (1994).

\bibitem{kappaL}
J. C. Brice, in Ref.~\protect\onlinecite{Brice}, Ch. 1.9.

\end{references}
\end{document}